\documentclass[conference]{IEEEtran}
\usepackage{graphicx} 
\usepackage{url}

\begin{document}
\title{Safety Factories - a Manifesto}

\IEEEoverridecommandlockouts
\IEEEaftertitletext{\vspace{-2\baselineskip}}
  
\author{\IEEEauthorblockN{Carmen Cârlan}
\IEEEauthorblockA{
\textit{TÜV SÜD GmbH, Germany}\\
carmen.carlan@tuvsud.com}
\and
\IEEEauthorblockN{Daniel Ratiu}
\IEEEauthorblockA{
\textit{CARIAD, Germany}\\
daniel.ratiu@cariad.technolgoy}
\and
\IEEEauthorblockN{Michael Wagner}
\IEEEauthorblockA{
\textit{Edge Case, USA}\\
mwagner@ecr.ai}
}
\maketitle

\begin{abstract}
Modern cyber-physical systems are operated by complex software that increasingly takes over safety-critical functions. 
Software enables rapid iterations and continuous delivery of new functionality that meets the ever-changing expectations of users. 
As high-speed development requires discipline, rigor, and automation, \emph{software factories} are used. These entail methods and tools used for software development, such as build systems and pipelines. 
%
%
%
To keep up with the rapid evolution of software, we need to bridge the disconnect in methods and tools between software development and safety engineering today. We need to invest more in formality upfront -- capturing safety work products in semantically rich models that are machine-processable, defining automatic consistency checks, and automating the generation of documentation -- to benefit later. 
Transferring best practices from software to safety engineering is worth exploring. We advocate for \emph{safety factories}, which integrate safety tooling and methods into software development pipelines. 
\end{abstract}


\section{Safety Engineering in the Era of Software-Defined-X}
\label{sec:introduction}


{\bf Software Defined X} More and more of today's cyber-physical systems are software-defined -- they feature a stable hardware platform and most of the functionality is realized in software. This enables high flexibility in development, early delivery of a minimum viable product, and continuous delivery of new functionality that meets the evolving needs of customers. 
%
Despite containing glitches and bugs, at their core, many of these systems exhibit a high level of dependability. 

At the core of implementing the continuous software engineering approach is a set of practices, methods, and tools generically known as \emph{software factory}. A software factory is a toolbox of methods that implement the "machinery" for continuous delivery of software and exhibit recurrent characteristics.
%
First, there is a strong \emph{focus on automation} -- building, analyzing, testing, and deploying the system is done in an automated fashion. 
Automation takes advantage of \emph{incrementality} -- whenever changes occur, the effort of running the pipelines is proportional to the size of the change, not the size of the system. Build steps are either completely avoided, or aggressive caching of intermediate build products of previous builds is used. Workflows are automated, and quality gates (e.g., test coverage, static analyses) are enforced.
Second, to enable automation, the content (code, tests, configuration, ...) has a \emph{high degree of formalization} -- it is captured in machine-processable form using semantically rich languages -- a paradigm often referred to as \emph{X-as-Code}. Having the configuration as code is essential for achieving reproducible builds: on every machine and at any state saved in the repository in the past, the entire system can be reconstructed by the push of a button. 
Third, there are \emph{no islands of content}; everything needed by a project is available in a mono-repository, which is visible to all project members. Redundancies are avoided by maintaining \emph{single-source-of-truth} -- each information is saved exactly once, and when views are needed, they are generated via pipelines. 
Fourth, there is a high degree of \emph{discipline and accountability} in the team given the possibility to track every change of every line of code. Every change is associated with a specific commit linked to pull requests that contain the set of verification steps performed, including who reviewed the code and what comments were implemented. Pull requests are themselves linked to a tracker that documents the intent of the change. Whereas code reviews are to be conducted by specific groups of individuals (code owners) responsible for the parts of the content that have been modified. 


{\bf Safety} is an emerging property; even if the functionality is primarily developed in software, safety can only be verified at the system level, encompassing all hardware and mechanical components. There is a legal obligation to ensure the safety of the entire product, and 
a safety process is executed parallel to the system development process. That is, specific safety work products are generated, including the system safety case.
Whenever a change occurs, safety standards require performing an impact analysis that describes the type of change, affected work products, and the potential impact on functional safety. Based on the results of the impact analysis, actions need to be taken to re-establish the validity of the system's safety case. 
%
In more strongly regulated industries, such as railways or aerospace, every software release, including changes that impact the system's safety case, must be preceded by an independent assessment.

The need and challenges of integrating safety engineering with agile and continuous delivery approaches have been described in several research papers \cite{2018_kasauli_safety_critical_systems_and_agile_development_mapping_study,2018_macgregor_challenges_in_assuring_highly_complex_high_volume_safety_critical_systems,2017_voest_keeping_continuous_deliveries_safe,2020_zeller_industrial-roadmap_for_continuous_delivery_for_safety_critical_software}.  
Other published research proposes process and methodical approaches for integrating safety engineering into continuous delivery pipelines and performing iterative safety analysis in parallel with system development  \cite{2019_warg_continuous_deployment_with_continuous_assurance_cases,2022_munk_safeops,2024_cassel_safedevops}.
%
We advocate the need to re-think the safety engineering \emph{methods and tools} to bring them closer to "software factories" -- build a \emph{"safety factory"}.


There are recurring symptoms that organizations exhibit that prevent a deep integration between safety and software processes. For example, safety tools are isolated from each other and from software engineering tools, each producing content with its own persistence (e.g., databases, different file formats). This creates islands of (partially overlapping) content and makes it challenging to automatically check the traceability and consistency rules between content produced by different tools. Consequently, currently, safety experts invest manual effort to maintain consistency between content islands. Further, the tools employ different configuration management strategies, which leads to challenges in establishing baselines and managing multiple development branches, as they span multiple tools. Fine-grained reproducible builds are hard to achieve. In addition, teams and roles (e.g., safety, system, and software engineers) are divided into tooling silos that do not communicate well with each other, resulting in a lack of transparency between work products and teams. Documentation is produced "just" to satisfy process and legal requirements, having an informal format, such as plain text.
And the information sent to assessors is document-based as opposed to having access to the development environment and being able to navigate and query the development system. 

\section{Safety Factories}
\label{sec:safety_factories}





The main ideas behind safety factories are high automation of the system and safety engineering processes, integration of the content produced by system, safety, and software engineers, and using machine-readable safety cases at the core of continuous system development pipelines as quality (i.e., safety) gates for the success of system builds. 
To implement a safety factory, the following principles shall be followed:

\textbf{1)} 
\emph{Treat Safety Work Products as Code.} Formalize all safety work products using domain-specific, semantically rich languages that allow machine processing and continuous execution of automated analyses. Store and manage safety work products, including the system safety case, alongside software code and tests, in the same repository, managed by a single version control system. To achieve this, the current practice of safety experts needs to change so that they start using machine-processable specifications for safety work products.




\textbf{2)} \emph{Single Source of Truth.} Each information is saved exactly once in the repository. When different views are needed, they are automatically generated. Having a single source of truth avoids repetition and inconsistencies. 

\textbf{3)} \emph{Automated Impact Analysis.} 
Rigorously specify the dependencies between development and safety work products. The dependencies should be detailed enough to identify all potentially impacted entities by a change, thereby avoiding false negatives, and yet high-level enough not to generate too many false positives. Use the system's safety case as a semantically enriched traceability mechanism, specifying how each work product is used in conjunction with other work products to fulfill the system's safety goals. 


\textbf{4)} \emph{Safety Builds.} Together with system builds, execute safety builds that continuously check the consistency of safety work products and the corresponding software and system artifacts. Furthermore, check whether the identified system safety goals are consistent with the system and if they are satisfied. 
The success of a system build would be demonstrated by a \textit{closed} safety case, whose claims are supported by sufficient evidence. 

\textbf{5)} \emph{The System Safety Case Drives the Safety Builds} 
Use system safety cases as a central place to specify constraints between different artifacts.
In addition, safety cases shall drive the continuous and incremental execution of safety engineering activities, addressing safety-relevant emergent behaviors and feature interactions. Such activities, especially verification and validation, shall be automated,
and manual reviews shall be conducted "just" as complementary verification methods.

\textbf{6)} \emph{Safety Case is Eventually Consistent} To ensure consistency and completeness of the safety case, we require strict and semantically rich automated checks. However, during system development, some of these checks may fail, for example, due to inevitable temporary unknowns that arise during system development. If temporary inconsistencies are present (e.g., due to a lack of knowledge at a specific point), they should be automatically detected, temporarily whitelisted, and continuously handled. The build is not "green" from the safety case perspective, but the gaps are temporarily tolerated and made transparent. Deciding which checks to specify to avoid unnecessary build failures is not a trivial task. Independent assessors might support this task. 
    

\textbf{7)} \emph{Live Documentation Close to the Content:} Replace plain text documentation with machine-interpretable work products, which can be automatically queried. Furthermore, replace "reading" the documentation with an interaction or conversation with the development environment. When needed, automatically generate documentation for and from safety work products. Assessors will work with live models that can be queried interactively as opposed to PDF documents.


\textbf{8)} \emph{Company-wide Safety Mindset.} To properly live a safety culture, each team member must be aware of the general safety goals and the expected contribution of his/her work to these goals. To this end, the safety case can serve as the single source of truth, used by each member of the project to understand their role and contribution to the safety outcome.

\textbf{9)} \emph{Accountability/Ownership as a Given} Modern version control systems for code can track fine-grained changes over the entire history of an artifact. We can take advantage of these features to keep track of accountability/ownership for each line of the safety and system artifacts in an automated fashion (who created it, who verified, who changed, based on whose request, when were all these steps done). \\

We are experimenting with operationalizing these principles in FASTEN - \url{https://sites.google.com/site/fastenroot/}\\

Edge Case is developing its nLoop live safety case platform in accordance with principles discussed here.


\bibliographystyle{amsplain}
\bibliography{bibliography.bib}

\providecommand{\bysame}{\leavevmode\hbox to3em{\hrulefill}\thinspace}
\providecommand{\MR}{\relax\ifhmode\unskip\space\fi MR }
\providecommand{\MRhref}[2]{%
  \href{http://www.ams.org/mathscinet-getitem?mr=#1}{#2}
}
\providecommand{\href}[2]{#2}
\begin{thebibliography}{1}

\bibitem{2024_cassel_safedevops}
A.~Cassel, F.~Beckman, B.~Haraldsson, M.~Törngren, D.~Sandberg, M.~Nyberg,
  M.~Hasselberg, S.~Holmqvist, M.~Erdogan, T.~Strandberg, R.~Johansson, and
  X.~Zhang, \emph{{DevOps Safety – Do Not Get Left at the Station!}},
  SAFECOMP Position Papers, 2024.

\bibitem{2018_kasauli_safety_critical_systems_and_agile_development_mapping_study}
R.~Kasauli, E.~Knauss, and B.~Kanagwa, \emph{Safety-critical systems and agile
  development: A mapping study}, SEAA, 2018.

\bibitem{2018_macgregor_challenges_in_assuring_highly_complex_high_volume_safety_critical_systems}
J.~MacGregor and S.~Burton, \emph{Challenges in assuring highly complex, high
  volume safety-critical software}, SASSUR, 2018.

\bibitem{2022_munk_safeops}
P.~Munk and M.~Schweizer, \emph{{DevOps and Safety? SafeOps!}}, SASSUR'22.

\bibitem{2017_voest_keeping_continuous_deliveries_safe}
S.~V\"{o}st and S.~Wagner, \emph{Keeping continuous deliveries safe}, ICSE'17.

\bibitem{2019_warg_continuous_deployment_with_continuous_assurance_cases}
F.~Warg, H.~Blom, J.~Borg, and R.~Johansson, \emph{Continuous deployment for
  dependable systems with continuous assurance cases}, WOSOCER, 2019.

\bibitem{2020_zeller_industrial-roadmap_for_continuous_delivery_for_safety_critical_software}
M.~Zeller, D.~Ratiu, M.~Rothfelder, and F.~Buschmann, \emph{{An Industrial
  Roadmap for Continuous Delivery of Software for Safety-critical Systems}},
  SAFECOMP Position Paper, 2020.

\end{thebibliography}

\end{document}